# Parametric Optomechanical Oscillations in Two-Dimensional Slot-Type High-*Q* Photonic Crystal Cavities

Jiangjun Zheng,[1,+,*] Ying Li,[1,+] Mehmet Sirin Aras,[1] Aaron Stein,[2] Ken L. Shepard,[3] and Chee Wei Wong,[1,*]

[1] Optical Nanostructures Laboratory, Columbia University, New York, NY 10027, USA

[2] Brookhaven National Laboratory, Upton, NY 11973, USA

[3] Department of Electrical Engineering, Columbia University, New York, NY 10027, USA

Abstract: We experimentally demonstrate an optomechanical cavity based on an air-slot photonic crystal cavity with optical quality factor $Q_o$=4.2×10$^4$ and a small modal volume of 0.05 cubic wavelengths. The optical mode is coupled with the in-plane mechanical modes with frequencies up to hundreds of MHz. The fundamental mechanical mode shows a frequency of 65 MHz and a mechanical quality factor of 376. The optical spring effect, optical damping and amplification are observed with a large experimental optomechanical coupling rate $g_{om}/2\pi$ of 154 GHz/nm, corresponding to a vacuum optomechanical coupling rate $g^*/2\pi$ of 707 kHz. With sub-mW or less input power levels, the cavity exhibits strong parametric oscillations. The phase noise of the photonic crystal optomechanical oscillator is also measured.

[+] Equal contribution
[*] Electronic addresses: jz2356@columbia.edu and cww2104@columbia.edu

Photonic crystal (PhC) slab is a versatile platform for optical device development and scientific observations [1]. By adding defects to photonic crystal slab or waveguide, ultra-small-volume ($V$) cavities can be engineered to have a ultra-high optical quality factor $Q$ [2]. The large $Q/V$ ratio indicates a strong internal optical field, which enables strong on-chip light-matter interactions due to the optical nonlinearities, such as free-carrier dispersion and two-photon absorption [3]. In recent years, the field of cavity optomechanics has attracted extensive attention for investigation of light-structure interactions [4-6]. Due to the improvement of micro- and nano-fabrication techniques, on-chip devices with unique geometries have been successfully demonstrated with well-coupled optical and mechanical modes, and much work has been done on actuation and cooling of their mechanical motions using optical pumps [7]. Compared with device platforms for microtoroid and microsphese etc., PhC slabs are shown to be an attractive, flexible platform for developing integrated devices that support well-coupled ultrasmall high-$Q$ optical modes and small resonant mechanical modes [8,9]. The strong internal optical field of high-$Q/V$ PhC cavities corresponds to a strong optical force with a low input optical power. The small size of the resonant mechanical device corresponds to a small physical mass or effective mechanical mass, which typically leads to a high mechanical frequency. The high mechanical frequency is important for reaching the sideband resolved regime, and is widely tunable by varying the geometrical parameters of the photonic crystal cavities [10]. The mechanical $Q$ is another important factor for optomechanical applications which is proportional to the coherence time of the mechanical vibrations. It is possible to improve the mechanical $Q$ of the PhC resonators by carefully engineering the anchor geometries to reduce the radiation losses. In particular, it is suggested that the radiation losses can be eliminated by anchoring the PhC resonator to artificial materials with complete mechanical band gaps covering the mechanical resonances [11-13]. Furthermore, the optomechanical coupling rates $g_{om}$ of the reported photonic crystal cavities are high with well-overlapped optical and mechanical fields [9,10,14]. Among these cavities, the differential in-plane mechanical modes of the air-slot photonic crystal cavities show relatively higher optomechanical coupling rates since the optical resonant frequencies are very sensitive to the slot-width variations [15,16]. In our previous work [15], we have theoretically investigated the dispersive optomechanical coupling in an air-slot mode-gap photonic crystal cavity. Here, we experimentally demonstrate the optomechanical properties of fabricated samples. With a large optomechanical coupling rate, the optical induced spring effect, damping and amplification of the mechanical modes are observed with measurements in air. Behaviors above the parametric instability threshold are shown. Finally, the phase noise of the optomechanical oscillator is analyzed.

The air-slot optomechanical cavity used in these experiments is similar in geometry to the one examined in our former work [15,17]. An air-slot is created along a regular mode-gap cavity [18] and holes in the cavity center are shifted by a few nanometers. The device is fabricated in silicon-on-insulator (SOI) substrate with a 250-nm thick silicon layer on top of a 3-$\mu$m thick buried oxide layer. Electron-beam lithography is used to define the device in a 300-nm thick resist (ZEP520A) coating. After developing, the pattern is transferred onto the silicon layer by using plasma etching. The remaining resist is removed by the PG solvent remover. Finally, the device is air-bridged by soaking it in a 5:1 buffered oxide etchant for 7 minutes to remove the sacrificial oxide under the device. A scanning electron microscope (SEM) image of the fabricated sample is shown in Fig. 1(a). After calibration, statistical analysis of the SEM image shows that the measured lattice period $a$ is 525 nm, radius $r$ is 0.375$a$ and slot width $s$ is 105 nm. The lattice holes at the cavity center are shifted by 15 nm, 10 nm and 5 nm, respectively.

The simplified measurement scheme is shown in Fig. 1(c). The light source is a high performance tunable laser (Santec TSL510 Type C, $\lambda$ = 1500–1630 nm). The polarization state of the input light is optimized by using a polarization controller (PC). The device under test (DUT) is probed by a single-mode fiber taper with a single loop created at its thinnest region. The diameter of the loop is around 70 $\mu$m. To tune the coupling strength, the lateral taper-cavity distance is adjusted gradually with a step resolution of 25 nm. The looped fiber taper is then loaded on one side of the device as shown in Fig. 1(b). It is necessary to anchor the taper to the device to avoid fiber-taper oscillations due to the optical force when the laser wavelength is swept across the optical resonances. The optical transmission is monitored by a fast detector (New Focus Model 1811) and a slow detector simultaneously. The radio-frequency (RF) signal from another fast detector (New Focus Model 1611-AC) is analyzed by a power spectrum analyzer (PSA, Agilent E4440A) and an oscilloscope. The phase noise of the optomechanical oscillator is measured by a phase noise analyzer (Agilent E5052B) which employs the cross-correlation technique.

Fig. 1(d) shows the optical transmission measurement with two cavity modes examined. The high-order mode is a relatively delocalized mode at 1575.49 nm. The fundamental mode with resonant wavelength of 1587.645 nm is the expected high-$Q$ mode based on the mode-gap effect. The measured loaded quality factor $Q_{tot}$ is 32,900 and the intrinsic quality factor $Q_{int}$ is 42,000. A three-dimensional finite-difference time-domain (FDTD) simulation [19], with the refractive index of silicon $n_{si}$ = 3.4, shows that the resonant wavelength for this mode is about 1584.95 nm, which matches our measurement very well. The theoretical intrinsic optical quality factor is higher than $1.3 \times 10^6$. The measured $Q_{int}$ is much less than the theoretical $Q$, and is limited by the fabrication quality and surface-state absorption [16]. The optical mode volume is about 0.051$(\lambda/n_{\text{air}})^3$. The electrical field intensity of the cavity mode is also shown in Fig. 1(d). The maximum field intensity is located at the cavity center in the air slot. Due to its field distribution feature, the fundamental optical mode is well coupled with the selected differential in-plane mechanical modes, which will be shown later.

Fig. 1(e) shows the measured RF power spectrum density (PSD) with incident power of 81 $\mu$W at the blue detuning of the fundamental optical mode. The resolution bandwidth is 10 kHz. Four major peaks are shown with frequencies of 65.68 MHz, 131.22 MHz, 196.93 MHz and 225.97 MHz. The first peak clearly indicates the fundamental mechanical mode, while the second and third peaks indicate its second and third harmonics respectively. The high-order harmonics reflect the nonlinear optomechanical transduction from the Lorentzian optical lineshape. The mechanical modes are identified by using the finite-element method (FEM). In the frequency range up to 320 MHz, over sixty mechanical modes are obtained which are classified into three kinds: in-plane, out-of-plane and twisting modes. Each kind is composed by mode pairs of differential and common motions. The mechanical modes detected by the optical method are the modes with relatively large coupling with the optical mode. The optomechanical coupling rates $g_{\text{om}} = d\omega_o/dx$ are inversely proportional to the optomechanical coupling lengths. For the Fabry–Pérot cavity, $g_{\text{om}}$ is easily determined by the cavity length. For complex cavity geometries such as the one examined here, $g_{om}$ is derived by using a perturbation theory for Maxwell's equations with shifting material boundaries and calculated with the non-perturbed optical and mechanical fields [20, 21]. For the slot-type cavity here, the resonant optical frequency is strongly dependent on the slot width. Thus, the differential in-plane mechanical modes with relatively larger displacements at the cavity center have larger optomechanical coupling rates. As is expected, two of these modes are founded at frequencies 64.82 MHz and 231.41 MHz, which match very well with the experimentally observed major peaks. These two in-plane mechanical modes exhibit modeled optomechanical coupling rates of $g_{om}/2\pi$ = 564 GHz/nm and 393 GHz/nm respectively, which are at least fifty times larger than the other mechanical modes. Their displacement fields are shown in Fig. 1(e). The fundamental differential in-plane mechanical mode has an effective mass of 6.11 pg and an effective mechanical volume of 2.62 $\mu$m$^3$. The vacuum optomechanical coupling rate $g^*/2\pi$ is 2.59 MHz which is large even for photonic crystal structures. In above simulations, density of silicon is 2.329×10$^3$ kg/m$^3$, Poisson's ratio is 0.28, and Young's modulus is 170 GPa [22].

The fundamental differential in-plane mechanical mode is selected for studying the optomechanical dynamics. Fig. 2(a) shows the optical transmission by the slow detector under different incident powers. As the power increases from 5.1 $\mu$W to 162.2 $\mu$W, the optical bistability arises mainly due to the two-photon absorption (TPA) [3]. The optical modification of the Brownian mechanical vibrations is measured by sweeping the incident wavelength across the resonance. As an example, Fig. 2(b) shows the obtained two-dimensional (2D) map of the RF PSD with the incident optical power of 51 $\mu$W. The wavelength is tuned with a 1 pm step. The RF PSD is given at each step. The blue detuning side and the red detuning side are separated by the zero-detuning line, where the RF signal is modulated in phase rather than in amplitude. The optical spring effect is shown with obvious increased mechanical frequency at the blue detuning side for this 65 MHz resonator. In the measurements, an incident power as low as 5.1 $\mu$W is able to introduce an observable mechanical frequency change due to the optical force. The 2D RF PSDs are analyzed and summarized for different incident powers in Fig. 2(c). The linewidth and frequency are extracted for the detuning with maximized optical spring effect. For powers more than 32.2 $\mu$W, there are bends on the curves for the red side detuning due to the difficulty of setting the laser wavelength to the theoretical optimal detuning. Using a linear fit of the non-bended parts of the curves, it is obtained that the intrinsic mechanical frequency $\Omega_m/2\pi$ is 64.99 MHz and the cold cavity mechanical quality factor $Q_m$ is 367. Furthermore, the experimental optomechanical coupling rate is also obtained with $g_{om}/2\pi = 154.1$ GHz/nm, correspond to a vacuum optomechanical coupling rate of $g^*/2\pi = 707.2$ kHz. The discrepancy between the modeled and experimental optomechanical coupling rates is believed to be due to the coupling with the adjacent flexural mechanical modes [16]. Fig. 2(d) shows the measured and fitted RF PSDs for a red detuning point (P1) and a blue detuning point (P2) as indicated in Fig. 2(c). For P1, the mechanical frequency decreases by 345 kHz, and the Brownian motion of this mechanical mode dampens with an effective $Q_m$ of 282.4. On the contrary, for P2 the mechanical frequency increases by 730 kHz, and the Brownian motion is amplified with a $Q_m$ of 1358.3. The above observations show that the picogram slot-type cavity has a large optomechanical coupling rate which is of key importance for optomechanical studies. In addition, the mechanical frequencies for devices with different lengths and PhC lattice constants vary between 50 MHz to 120 MHz in the experiments.

With a proper high input power, the optomechanical cavity starts to oscillate periodically when the intrinsic mechanical energy dissipation is overcome by the optical amplification. Fig. 3(a) shows the optical transmission for an input power of 520 $\mu$W by use of an oscilloscope. The blue line is obtained by the fast detector (New Focus Model 1811), and the red curve is obtained by using a low-pass filter with a cutoff frequency of 1.9MHz. The optically-driven oscillations rise suddenly near the wavelength 1587.65 nm which indicates the above-threshold behavior [23]. The filtered signal shows a sudden drop at that wavelength as seen in Fig. 3(a). In Fig. 2(c), the trend of linewidth for the blue detuning indicates a threshold power slightly higher than 100 $\mu$W with optimized detuning. This is confirmed by a series of swept-wavelength measurements with different incident powers. In Fig. 3(b) and (c), the transmission oscillations in frequency and time domains are given by using a low-noise fast detector with 1 GHz bandwidth (New Focus Model 1611-AC). It is shown that the detuning gets smaller, the second harmonic get relatively stronger and a second peak is rising in a single oscillation period. It indicates that the resonant optical frequency shift induced by optical pump gets comparable to the optical resonance linewidth. The effective real-time resonant wavelength is smaller than the incident wavelength for a certain time in a period, which is shown by the normalized real-time detuning in Fig. 3(c). By using a higher incident power, the nonlinearity of the optical power oscillations will get stronger due to a larger amplitude of the sinusoidal mechanical vibrations. An example with 1.3-mW input power is shown in the bottom panel of Fig. 3(b). Harmonics with frequency up to 1 GHz are observed. The lineshape and spectrum show the dynamic interactions of the internal optical cavity field with the mechanical movement for different detuning and incident power, which is accurately described by temporal coupled-mode theory [15,24].

With the above-threshold condition satisfied, the optomechanical cavity is viewed as a self-sustained close-loop oscillator which can be utilized as a photonic clock. The amplitude noise is suppressed above the threshold due to the inherent amplitude-limiting mechanism present in harmonic oscillators [25]. The phase noise in the system is measured with phase noise analyzers for high-precision estimation of the short-term linewidth and can be explained with Lesson's model [26]. In this model, there is always a $1/f^3$ region at small frequency offsets due to $1/f$ noise. After the corner frequency is the $1/f^2$ region due to the white frequency noise which reflects the measurement limitations [27]. Here, the phase noise analyzer (Agilent E5052B) employs a two-channel cross-correlation technique for accurately analysis of the RF signal carried by the transmission light. In Fig.4, the typical phase noise versus offset frequency is obtained with an input laser power of 1.3 mW. By fitting with a piecewise function following the Lesson's model, it shows that the corner frequency is 25.4 kHz. For offset frequencies less than 25.4 kHz, the $1/f^3$ dependence is predominantly related to relatively slow environmental vibrations. For offset frequencies larger than 25.4 kHz, the $1/f^2$ dependence is related to intrinsic properties of the oscillator. We note that the fitting for the $1/f^2$ region shows an averaged dependence rather than a well-matched dependence. The averaged dependence is possibly due to the noise contamination by coupling to nearby low-$Q$ mechanical modes since the feedback loop of optomechanical oscillators is inherent. The anchored fiber taper also adds noise to the system [27]. Assuming a 1MHz offset point in the $1/f^2$ regime, the phase noise of 117 dBc/Hz there corresponds to a short-term linewidth of 12.5 Hz. The root-mean-square (RMS) timing jitter integrated from 1 kHz to 1 MHz offset frequency is 758 ps with major contributions from 1 kHz to 10 kHz. The RMS timing jitter from 10 kHz to 1 MHz is about 81 ps. In future experiments, the phase noise performance of the current optomechanical oscillator will be improved by further engineering the mechanical $Q$ and taking measurements in a more stable environment such as a vacuum chamber.

In conclusion, we have experimentally demonstrated the optomechanical behaviors of a slot-type high-optical-$Q$ PhC cavity in air at room temperature. The optical and mechanical simulations match the experiments well. The optomechanical coupling rate between the selected mechanical modes and optical modes is quantified at $g_{om}/2\pi = 154$ GHz/nm. With the large optomechanical coupling, the optical spring effect, optical damping and amplification of the mechanical mode are all clearly exhibited. In the above threshold regime, time-domain oscillations are observed and the phase noises are analyzed. The device is shown to be a promising candidate for studying optomechanical dynamics and applications. Further experimental study will focus on improving the optical $Q$ by optimizing the fabrication process and the mechanical $Q_m$ by modifying the anchor geometry [28, 29]. The measured mechanical frequency is widely tunable by varying geometrical parameters. Additionally, it has the potential to be improved for more challenging tasks such as side-band resolved cooling of the mechanical modes and work as a low-power low-noise photon clock.

The authors acknowledge discussions with Hong X. Tang, J. Gao, and J. F. McMillan on the optomechanical studies, and with Sunil Bhave and Harish Krishnaswamy on the phase noise measurements. This work is supported by Defense Advanced Research Projects Agency (DARPA) DSO with program manager Dr. J. R. Abo-Shaeer under contract number C11L10831. Device fabrication is carried out in part at the Center for Functional Nanomaterials, Brookhaven National Laboratory, which is supported by the U.S. Department of Energy, Office of Basic Energy Sciences, under Contract No. DE-AC02-98CH10886.

References:

[1]. S. Noda, A. Chutinan, and M. Imada, Nature **407**, 608 (2000).

[2]. M. Notomi, E. Kuramochi, and T. Tanabe, Nat. Photonics **2**, 741 (2008).

[3]. X. Yang, C. Husko, and C. W. Wong, Appl. Phys. Lett. **91**, 051113 (2007); C. Husko, A. Rossi, S. Combrie, Q. V. Tran, F. Raineri, and C. W Wong, Appl. Phys. Lett. **94**, 021111 (2009).

[4]. T. J. Kippenberg, and K. J. Vahala, Opt. Express **15**, 17172 (2007).

[5]. M. Li, W. Pernice, C. Xiong, T. Baehr-Jones, M. Hochberg, and H. Tang, Nature **456**, 480 (2008).

[6]. D. Van Thourhout, and J. Roels, Nat. Photonics **4**, 211 (2010).

[7]. A. Schliesser, R. Rivière, G. Anetsberger, O. Arcizet, and T. J. Kippenberg, Nat. Physics **4**, 415 (2008).

[8]. Y.-G. Roh, T. Tanabe, A. Shinya, H. Taniyama, E. Kuramochi, S. Matsuo, T. Sato, and M. Notomi, Physical Review B **81**, 121101R (2010).

[9]. M. Eichenfield, R. Camacho, J. Chan, Kerry J. Vahala, and O. Painter, Nature **459**, 550 (2009).

[10]. M. Eichenfield, J. Chan, R. M. Camacho, K. J. Vahala, and O. Painter, Nature **462**, 78 (2009).

[11]. S. Mohammadi, A. A. Eftekhar, A. Khelif, and A. Adibi, Opt. Express **18**, 9164 (2010).

[12]. Y. Pennec, B. Djafari Rouhani, E. H. EI Boudouti, C. Li, Y. EI Hassouani, J. O. Casseur, N. Papanikolaou, S. Benchabane, V. Laude, and A. Martinez, Opt. Express **18**, 14301 (2010).

[13]. A. H. Safavi-Naeini, and O. Painter, Opt. Express **18**, 14926 (2010).

[14]. E. Gavartin, R. Braive, I. Sagnes, O. Arcizet, A. Beveratos, T. J. Kippenberg, and I. Robert-Philip, Phys. Rev. Lett. **106**, 203902 (2011).

[15]. Y. Li, J. Zheng, J. Gao, J. Shu, M. S. Aras, and C. W. Wong, Opt. Express **18**, 23844 (2010).

[16]. A. H. Safavi-Naeini, T. P. M. Alegre, M. Winger, and O. Painter, Appl. Phys. Lett. **97**, 181106 (2010).

[17]. J. Gao, J. F. McMillan, M.-C. Wu, J. Zheng, S. Assefa, and C. W. Wong, Appl. Phys. Lett. **96**, 051123 (2010).

[18]. E. Kuramochi, M. Notomi, S. Mitsugi, A. Shinya, T. Tanabe, and T. Watanabe, Appl. Phys. Lett. **88**, 041112 (2006).

[19]. A. F. Oskooi, D. Roundy, M. Ibanescu, P. Bermel, J. D. Joannopoulos, and S. G. Johnson, Computer Physics Communications **181**, 687-702 (2010).

[20]. S. G. Johnson, M. Ibanescu, M. A. Skorobogatiy, O. Weisberg, J. D. Joannopoulos, Phys. Rev. E **65**, 066611 (2002).

[21]. C. W. Wong, P. T. Rakich, S. G. Johnson, M. Qi, H. I. Smith, E. P. Ippen, L. C. Kimerling, Y. Jeon, G. Barbastathis, and S.-G. Kim, Appl. Phys. Lett. **84**, 1242 (2004)

[22]. M. A. Hopcroft, W. D. Nix, and T. W. Kenny, Journal of Microelectromechanical systems **19**, 229 (2010).

[23]. Q. lin, J. Rosenberg, X. Jiang, K. J. Vahala, and O. Painter, Phys. Rev. Lett. **103**, 103601 (2009).

[24]. T. Carmon, H. Rokhsari, L. Yang, T. J. Kippenberg, and K. J. Vahala, Phys. Rev. Lett. **94**, 223902 (2005).

[25]. M. Hossein-Zadeh, H. Rokhasari, A. Hajimiri, and K. J. Vahala, Phys. Rev. A **74**, 023813 (2006). H. Rokhsari, M. Hossein-Zadeh, A. Hajimiri, and K. J. Vahala, Appl. Phys. Lett. **89**, 261109 (2006).

[26]. A. Hajimiri, and T. H. Lee, Kluwer Academic Publishers, 2003.

[27]. S.Tallur, S. Sridaran, and S. A. Bhave, Opt. Express **19**, 24522 (2011).

[28]. K. Wang, A.-C. Wong, and C. T.-C. Nguyen, J. of Microelectromechanical systems 9, 347 (2000).

[29]. G. Anetsberger, R. Rivière, A. Schliesser, O. Arcizet and T.J. Kippenberg, Nat. Photonics 2, 627 (2008).

Figures.

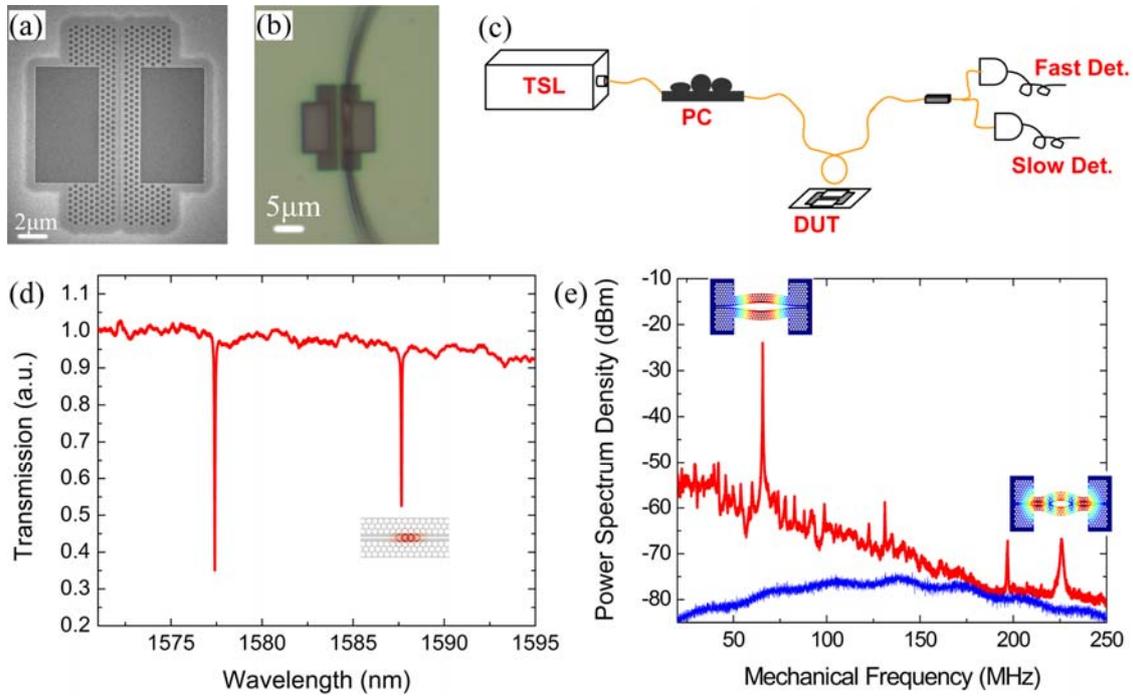

FIG. 1. (Color online) (a) SEM image of the fabricated sample. (b) Optical microscope image showing the fiber taper loop loaded on the right side of the cavity. (c) Simplified scheme of the measurements. (d) Normalized optical transmission by laser scanning showing two cavity modes. The inset below the fundamental optical mode shows its electrical-field energy distribution. (e) Red curve shows RF PSD at a blue detuning for the frequency range from 20 MHz to 250 MHz . The first three major peaks are related to the fundamental differential in-plane mechanical mode. The last major peak shows a high-order differential in-plane mechanical mode. FEM simulated frequencies matches the measured ones well. The displacement field distributions for these two modes are shown as insets on top of their RF peaks. The blue curve shows the measurement noise floor. The incident optical power is 81 $\mu$W and the resolution bandwidth is 10 kHz.

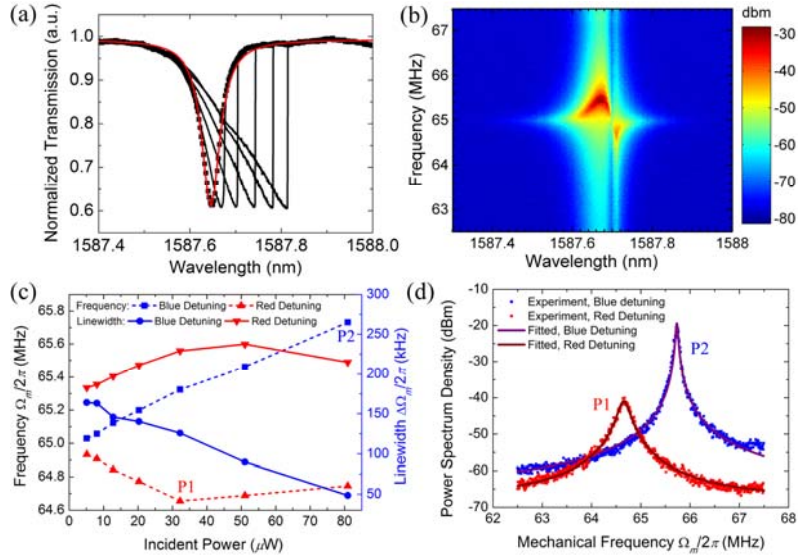

FIG. 2. (Color online) (a) Normalized optical transmission versus wavelength for increasing incident powers. Black dots for an incident power of 5.1 $\mu$W, black curves for 32.2 $\mu$W, 64.9 $\mu$W, 102.6 $\mu$W, 128.2 $\mu$W, and 162.2 $\mu$W respectively. The red curve is the fitted curve which gives a total quality factor $Q_{tot}$ = 32,900 and an intrinsic quality factor $Q_{int}$ = 42,000 under a low incident power. For higher incident powers, the red-shifted sharp edges indicate the thermal optical bistability. (b) Example RF PSD by sweeping the laser wavelength. (c) Mechanical frequency and linewidth versus incident power at the optimal detuning for the optical spring effect. (d) RF PSD for points P1 and P2 indicated in panel (c). The dots are the experimental values and the solid curves are the fitted.

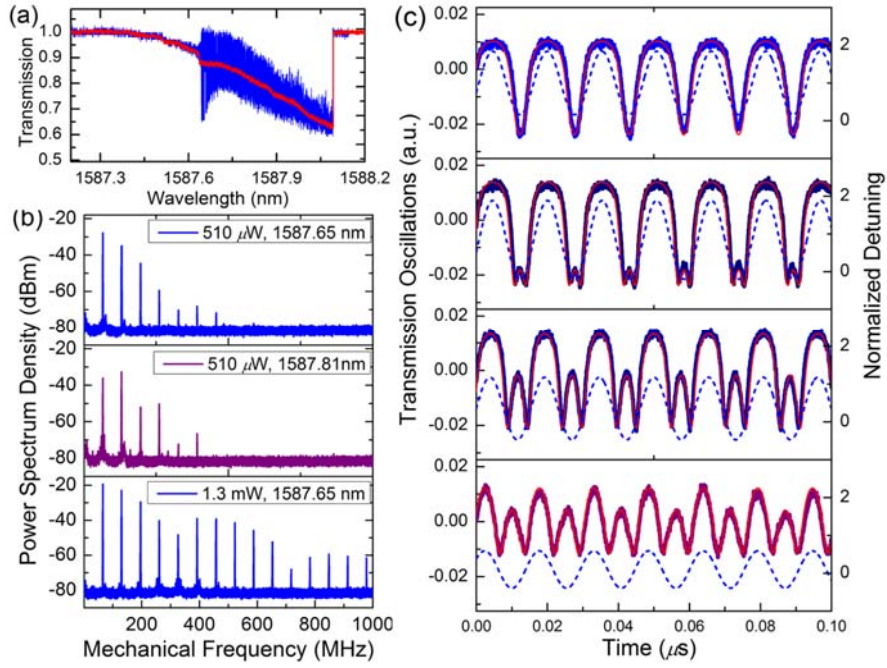

FIG. 3. (Color online) (a) Normalized optical transmission for an input power of 510 $\mu$W by sweeping laser wavelength across the optical resonance. Strong oscillations are shown by the blue curve, overlaid with the red low-pass-filtered signal. (b) Power spectrum density of the transmission oscillations for different incident powers (510 $\mu$W and 1.3 mW) and wavelengths (1587.65 nm and 1587.81 nm). The reference bandwidth is 100 kHz. (d) Transmission oscillations overlaid with modeled oscillations (Red curves) in time domain for incident power of 510 $\mu$W as used in (a). The incident wavelengths are varying from 1587.65 nm (the top panel) to 1587.81 nm (the bottom curve). The blue dashed curves show the corresponding real-time detuning. The detuning is normalized by the linewidth of the optical linewidth.

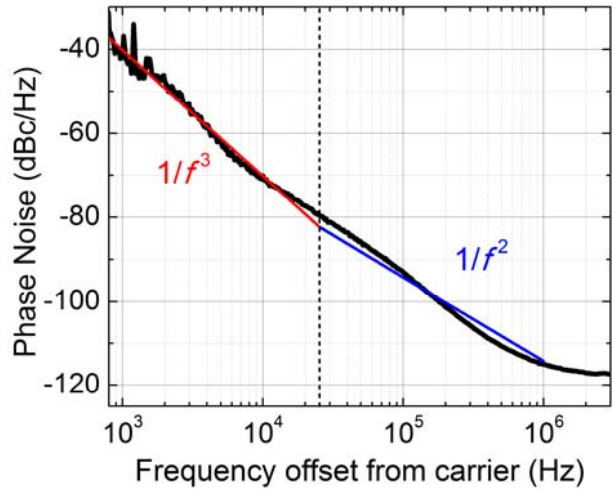

FIG. 4. (Color online) Example phase noise spectrum of the optomechanical oscillations obtained with input power of 1.3 mW. The black curve shows experimental data which is fitted by using a piecewise function. The red line indicates the $1/f^3$ dependence and the blue line indicates the $1/f^2$ dependence.